\documentclass{aa}
\usepackage[dvips]{graphics}
\begin{document}
\newcommand{\apj}[1]{{ApJ }{ #1}}
\newcommand{\apjs}[1]{{ApJS }{ #1}}
\newcommand{\aj}[1]{{AJ }{ #1}}
\newcommand{\aea}[1]{{A\&A }{ #1}}
\newcommand{\aeass}[1]{{A\&AS }{ #1}}
\newcommand{\mnras}[1]{{MNRAS }{ #1}}
\newcommand{\acas}[1]{{Acta Astron.\ }{ #1}}
\newcommand{\pasp}[1]{{PASP }{ #1}}
\newcommand{\pasj}[1]{{Publ.\ Astron.\ Soc.\ Japan\ }{ #1}}
\newcommand{\araa}[1]{{ARA\&A }{ #1}}
\newcommand{\apess}[1]{{Ap\&SS }{ #1}}
\newcommand{\nat}[1]{{Nat }{ #1}}
\newcommand{\sci}[1]{{Sci }{ #1}}
\newcommand{\kms}{$\rm{km\,s^{-1}}$} 
\newcommand{\mic}{$\mu$m}
\newcommand{\su}{$\surd$}
\newcommand{\ang}{$\rm \AA$\,}
\newcommand{\halfa}{H$\alpha$}

\def\ga{\mathrel{\mathchoice
{\vcenter{\offinterlineskip\halign{\hfil$\displaystyle##$\hfil\cr>\cr\sim\cr}}}
{\vcenter{\offinterlineskip\halign{\hfil$\textstyle##$\hfil\cr>\cr\sim\cr}}}
{\vcenter{\offinterlineskip\halign{\hfil$\scriptstyle##$\hfil\cr
>\cr\sim\cr}}}
{\vcenter{\offinterlineskip\halign{\hfil$\scriptscriptstyle##$\hfil\cr>\cr\sim\cr}}}}}

\def\la{\mathrel{\mathchoice
{\vcenter{\offinterlineskip\halign{\hfil$\displaystyle##$\hfil\cr<\cr\sim\cr}}}
{\vcenter{\offinterlineskip\halign{\hfil$\textstyle##$\hfil\cr<\cr\sim\cr}}}
{\vcenter{\offinterlineskip\halign{\hfil$\scriptstyle##$\hfil\cr
<\cr\sim\cr}}}{\vcenter{\offinterlineskip\halign{\hfil$\scriptscriptstyle##$\hfil\cr><cr\sim\cr}}}}}
 
\title{ISO spectroscopy of circumstellar dust in 14 Herbig Ae/Be systems: 
towards an understanding of dust processing.\thanks{Based on observations 
with ISO, an ESA project with instruments funded by ESA Member States 
(especially the PI countries: France, Germany, the Netherlands and the 
United Kingdom) and with the participation of ISAS and NASA.}}

\author{ 
G. Meeus\inst{1} \and 
L.B.F.M. Waters \inst{2,1} \and 
J. Bouwman \inst{2} \and 
M.E. van den Ancker \inst{2,3} \and 
C. Waelkens\inst{1}  \and 
K. Malfait\inst{1}
}

\institute{Astronomical Institute, KULeuven, Celestijnenlaan 200B,
B-3001 Heverlee, Belgium 
\and Astronomical Institute ''Anton Pannekoek'', University of Amsterdam,
Kruislaan 403, NL-1098 SJ Amsterdam, The Netherlands 
\and Harvard-Smithsonian Center for Astrophysics, 60 Garden Street, MS 42, 
Cambridge, MA 02138, USA }

\thesaurus{08(08.03.4; 08.16.5; 13.09.4; 07.19.1)}

\date{9 August 2000; 20 October 2000}
\offprints{G. Meeus \\ (e-mail: gwendolyn@ster.kuleuven.ac.be)}
\titlerunning{ISO spectroscopy of 14 Herbig Ae/Be stars}
\authorrunning{G. Meeus et al.} 
\maketitle
                               
%%%%%%%%%%%%%%%%%%%%%%%%%%%%%%%%%%%%%%%%%%%%%%%
\begin{abstract}

We present Infrared Space Observatory (ISO) spectra of fourteen
isolated Herbig Ae/Be (HAEBE) stars, to study the characteristics of
their circumstellar dust. These spectra show large star-to-star
differences, in the emission features of both carbon-rich and
oxygen-rich dust grains. The IR spectra were combined with photometric
data ranging from the UV through the optical into the sub-mm
region. We defined two key groups, based upon the spectral shape of the 
infrared region. The derived results can be summarized as follows:
(1) the continuum of the IR to sub-mm region of all stars can be 
reconstructed by the sum of a power-law and a cool component, which can 
be represented by a black body. Possible locations for these components 
are an optically thick, geometrically thin disc (power-law component) and 
an optically thin flared region (black body); 
(2) all stars have a substantial amount of cold dust around them, 
independent of the amount of mid-IR excess they show; 
(3) also the near-IR excess is unrelated to the mid-IR excess, indicating 
different composition/location of the emitting material; 
(4) remarkably, some sources lack the silicate bands; 
(5) apart from amorphous silicates, we find evidence for crystalline 
silicates in several stars, some of which are new detections; 
(6) PAH bands are present in at least 50\% of our sample, and their 
appearance is slightly different from PAHs in the ISM;
(7) PAH bands are, with one exception, not present in sources which only 
show a power-law continuum in the IR; their presence is unrelated to the 
presence of the silicate bands; 
(8) the dust in HAEBE stars shows strong evidence for coagulation; this dust 
processing is unrelated to any of the central star properties (such as age, 
spectral type and activity).

\end{abstract}

\section{Introduction}

\begin{figure*}
\resizebox{\hsize}{!}{{\rotatebox{90}{\includegraphics{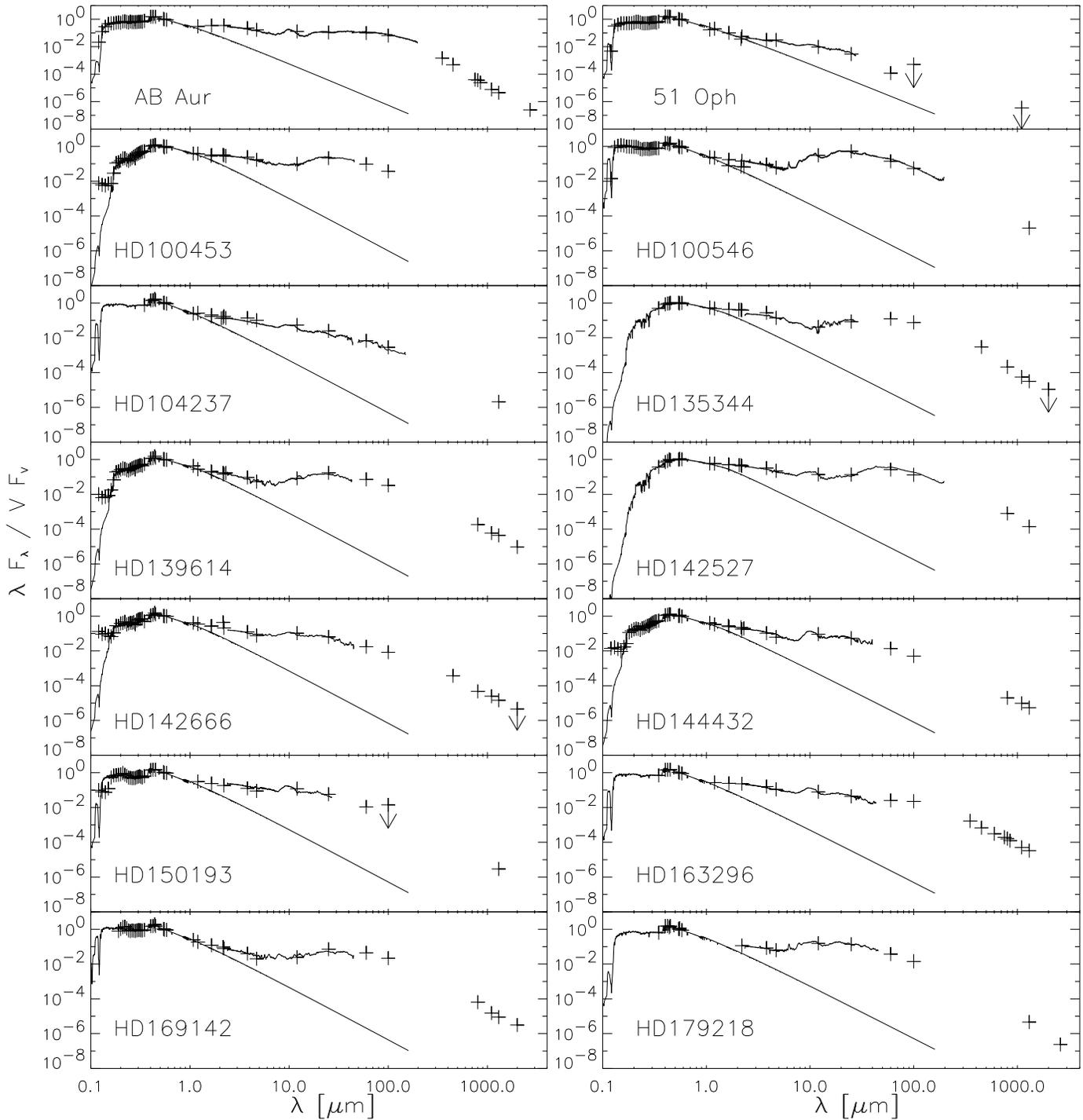}}}}
\vspace{-0.02cm}
\caption{ISO spectra of the 14 sample stars, superimposed on their spectral 
energy distributions. Crosses: observations; full line through the optical 
data: Kurucz model atmosphere; other full line: ISO-SWS/LWS observations; 
arrows indicate upper limits. The data are normalized to the V band.}
\label{allseds}
\end{figure*} 

\begin{figure*}
\resizebox{\hsize}{!}{{\rotatebox{90}{\includegraphics{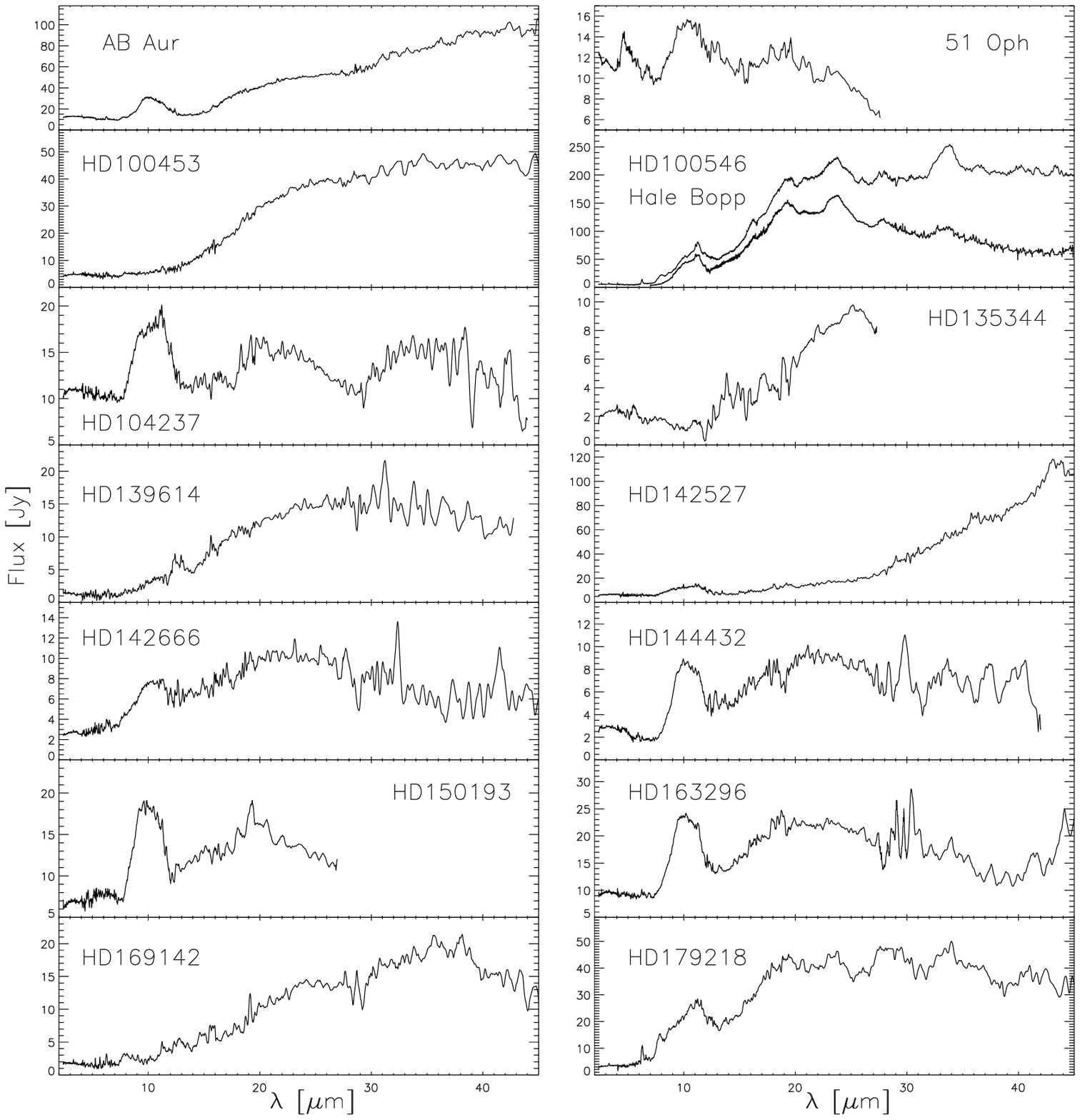}}}}
\caption{The ISO-SWS spectra of our programme stars. Together with HD100546, we 
also show the spectrum of comet Hale-Bopp (Crovisier et al. 1997) for 
comparison.}
\label{allspectra}
\end{figure*} 

A circumstellar (CS) disc is expected to be a natural byproduct of the
star forming process (e.g. Shu et al. 1987). This theoretical
expectation has obtained wide support from optical (e.g. McCaughrean
\& O'Dell 1996), infrared (e.g. Marsh et al. 1995) and millimetre
observations of young stars (e.g. Mannings \& Sargent 1997). The CS
disc is expected and observed to gradually disappear, but remnants are
still found around several Main-Sequence (MS) stars, such as Vega
(Aumann et al. 1984). Earlier and recent modelling of T Tauri discs
have shown that the most successful models are flaring passive discs
(Kenyon \& Hartmann 1987, Chiang \& Goldreich 1997).

Herbig Ae/Be stars (hereafter HAEBEs), first described as a group by
Herbig (1960), are believed to be the more massive analogues of T Tauri
stars. They are seen as the progenitors of Vega-type stars (for recent
reviews, see Waters \& Waelkens 1998; Natta et al. 1998). They are
characterized by large IR excesses due to thermal re-emission of CS
dust, show emission lines in their spectrum due to CS gas and have
masses between 2 and 8 $M_{\odot}$ (Herbig 1960). Infrared
spectroscopy offers a unique opportunity to scrutinize the composition
and characteristics of their CS dust. Recent ISO (Kessler et al. 1996)
studies have revealed a large variety in the properties of the dust
around HAEBEs, from which it became clear that their dust is
significantly different from that in the interstellar medium (Waelkens
et al. 1996, Malfait et al. 1998a, Malfait et al. 1999b, van den
Ancker et al. 1999).

\begin{center}
\begin{tiny}
\begin{table*}
\caption[]{Parameters of the fourteen programme stars. The groups are 
defined below (see Sect. 2.1).}
\label{sterren}
\begin{tabular}{cllrcrrcrcr}
\hline
\hline  \noalign{\smallskip}
     &       &(1)      &(2)                &(3)      &(4)&(5)&(6)&(7)&(8)&(9)\\
Group&Object & Spectral&$\mathrm{T_{eff}}$ & log $g$ &d  &log (Age)
&$\mathrm{E[B-V]_{CS}}$&$\mathrm{F_{800 \mu m} d^{2}}$ & $\mathrm{\lambda_{onset}}$ 
& $\mathrm{L_{IR}/L_{*}}$\\
&&Type&(K)&&(pc)&(yr)&&(mJy $\mathrm{pc^{2}}$)&($\mu$m)&\\[1ex]
\hline
\hline
\rule[3mm]{0mm}{1mm}
  &AB Aur     &B9/A0Ve&9750 &5.0&144&6.3&0.07      &11  $10^{6}$&1.1&0.48\\
Ia&HD100546   &B9Ve   &11000&4.5&103&$>$ 7.0&0.00  &12  $10^{6}$&1.2&0.51\\
  &HD142527   &F7IIIe &6250 &4.0&200&5.0&0.08      &166 $10^{6}$&1.1&1.06\\
  &HD179218   &B9e    &10000&5.0&240&5.0&0.06      &11  $10^{6}$&1.7&0.62\\[1ex]
\hline
\rule[3mm]{0mm}{1mm}
  &HD100453   &A9Ve   &7500 &4.5&-&-&0.06           &-          &1.2&0.54\\
Ib&HD135344   &F4Ve   &6750 &4.5&84&-&0.00          &4  $10^{6}$&1.1&0.44\\
  &HD139614   &A7Ve   &8000 &4.5&151&-&0.01         &14 $10^{6}$&1.5&0.39\\
  &HD169142   &A5Ve   &10500&4.5&145&-&0.00         &12 $10^{6}$&1.6&0.10\\[1ex]
\hline
\rule[3mm]{0mm}{1mm}
    &HD104237  &A4Ve  &10500&4.5&116&6.3&0.25    &3   $10^{6}$&1.2&0.13\\
    &HD142666  &A8Ve  &8500 &4.5&116&-&0.40      &4   $10^{6}$&1.1&0.28\\
IIa &HD144432  &A9Ve  &8000 &4.5&$>$ 200&-&0.05  &$>$ 4 $10^{6}$&1.1&0.26\\
    &HD150193  &A1Ve  &10000&4.0&150&$>$ 6.3&0.30&3   $10^{6}$&1.2&0.15\\
    &HD163296  &A3Ve  &10500&4.0&122&6.6&0.02    &34  $10^{6}$&1.2&0.16\\
    &51 Oph    &A0Ve  &10000&4.0&131&5.5&0.03    &$<$.8 $10^{6}$&2.3&$<$ 0.024\\[1ex]
\hline
\end{tabular} 
\footnotesize\\
\smallskip\\
References: (1): Malfait et al. (1998b), Dunkin et al. (1997), Gray \& Corbally 
(1998); (2), (3) and (6): Malfait et al. (1998b); (4) and (5): van 
den Ancker et al. (1999); (7) and (9): this study, based upon sub-mm
measurements by Sylvester et al. (1996), Mannings \& Sargent (1997), Walker 
\& Butner (1995), Henning et al. (1998) and Mannings \& Sargent (2000); (8):
this study.
\end{table*}
\end{tiny}
\end{center}

This paper is one in a series of papers based upon ISO-SWS
observations of HAEBE stars. In this study, we compiled a set of data
which include, next to the ISO spectra, also UV, optical, IR and
sub-mm photometry of a large sample of isolated HAEBE stars.  A
similar study was already presented by Sylvester et al. (1996) for a
sample of Vega-like systems. Their ground-based observations in the IR
with UKIRT are restricted to 2 ranges: 7.5-13.5 \mic~and 15.8-23.9
\mic.  Some of their sources (HD135344, HD139614, HD142666, HD144432, 
HD169142 and 51 Oph) are also part of our HAEBE sample, and it is interesting
to compare their results with ours.  In this paper we give an overview
of the IR features in our sample, together with a description of the
Spectral Energy Distributions (SED) and we propose a global model to
explain the SEDs.  In Sect. 2, we describe our sample stars and
their observations. We also present the SEDs (see Fig.~\ref{allseds})
and indicate observational trends. ISO-SWS spectra and an inventory of
solid state and PAH bands are shown in Sect. 3, where the individual
sources are discussed as well. In Sect. 4 we propose a global model,
and discuss grain processing. Our conclusions are summarized in
Sect. 5. In a forthcoming paper, detailed radiative transfer models
of some of the sources will be presented (Bouwman, Meeus \& Waters, in
preparation).

%%%%%%%%%%%%%%%%%%%%%%%%%%%%%%%%%%%%%%%%%%%%%%%%%%%%
\section{Targets and Observations}

The objects we selected are fourteen so-called isolated HAEBE stars.
These HAEBE stars are not located inside a star-forming region, but
show all the other characteristics of a HAEBE star and are presumably
the somewhat more evolved members of the HAEBE group. These objects
are best suited for our purpose, which is to discuss the evolution of
the CS disc, and offer the additional advantage that the spectra are
not strongly affected by loosely bound remnants of the star
formation process, but only show the emission of the disc. In
Table~\ref{sterren} we list the programme stars and among others their
main parameters: spectral type, effective temperature and log $g$.

The sources have been observed with the ISO Short Wavelength
Spectrometer (SWS; de Graauw et al. 1996) in mode AOT1. The spectra
cover an interval from 2 to 45~\mic. Some stars have also been
observed with ISO-LWS (Clegg et al. 1996), which covers a range from
45 to 200~\mic. These spectra were discussed by Malfait et al. (1998a,
1999a,b). In this study we will concentrate on the SWS data. The spectra
were reduced in a standard way using the ISO-SWS Interactive Analysis
(IA) tool containing pipeline processing steps of OLP version 8.5, and 
the ISO Spectral Analysis Package (ISAP version 1.6a). 
In Fig.~\ref{allspectra} the reduced SWS spectra are shown. For some sources
(HD135344, HD150193 and 51 Oph), the ratio signal-to-noise is so low at 
longer wavelengths that we had to leave out the part longwards of 28~\mic.

\subsection{Spectral Energy Distributions}

We also collected photometric data in the literature (Malfait et
al. 1998b and references therein; sub-mm data from Sylvester et
al. 1996, Mannings \& Sargent 1997, 2000; Walker \& Butner 1995,
and Henning et al. 1998), and composed for each star an SED, ranging
from the UV until the sub-mm region. In Fig.~\ref{allseds}, we show
for each of the fourteen sources its SED, combined with their
respective ISO spectrum. An appropriate Kurucz (1993) model atmosphere
was fitted through the optical data, representing the photospherical
contribution; it emphasizes the shape and the amount of the excess in
the IR and sub-mm region. The shortest wavelength at which an excess
is discernible is listed in Table~\ref{sterren} as
$\mathrm{\lambda_{onset}}$, with an uncertainty of 0.2~\mic. Also
shown in Table~\ref{sterren} is the derived fractional luminosity of
the dust, $\mathrm{L_{IR}/L_{*}}$, which is the ratio of the energy
radiated by the dust to the stellar luminosity. This ratio was
calculated as follows: first, we converted the data into the
$\mathrm{F_{\lambda}}$ versus $\lambda$ scale. Then we integrated the
Kurucz model over its entire wavelength range to calculate
$\mathrm{L_{*}}$. To obtain $\mathrm{L_{IR}}$, we first subtracted the
Kurucz model from the observations, and then integrated this curve
longwards of $\mathrm{\lambda_{onset}}$. Sylvester et al. (1996) have
already calculated this ratio for six of our sample stars, and their
results agree very well with four of our stars, while they agree 
less well for HD135344 (0.64) and HD1444432 (0.48). The values we 
obtained for $\mathrm{L_{IR}/L_{*}}$ are consistent with a passive 
reprocessing disc, except for the star HD142527 
($\mathrm{L_{IR}/L_{*}}$ = 1.06).

The dust continuum behaves very differently from source to source,
especially in the mid-IR (15 to 45~\mic). In some stars it is rising,
in other stars it is rather flat or even descending. Also the strength
of the dust continuum in these objects is very diverse: the
12~\mic~excess ranges between 3.5 and 7 magnitudes, the 60~\mic~excess
ranges between 4.5 and 12 magnitudes and the 1.3 mm excess between 10
and 13 magnitudes.
%[12]ex: 6.942 tot 3.535
%[25]ex: gaat van 10.482 tot 5.187
%[60]ex: 11.787 tot 4.452
%[1300]ex: 12.915 tot 10.012
A second important observational fact is the strong variation of the
strength of the 10~\mic~silicate feature from one object to
another. In some objects (e.g. HD144432) this feature is
very strong, in others (e.g.  51 Oph) it is less so,
and in several objects (e.g. HD169142) it is even absent.

\begin{figure}
\resizebox{\hsize}{!}{{\rotatebox{90}{\includegraphics{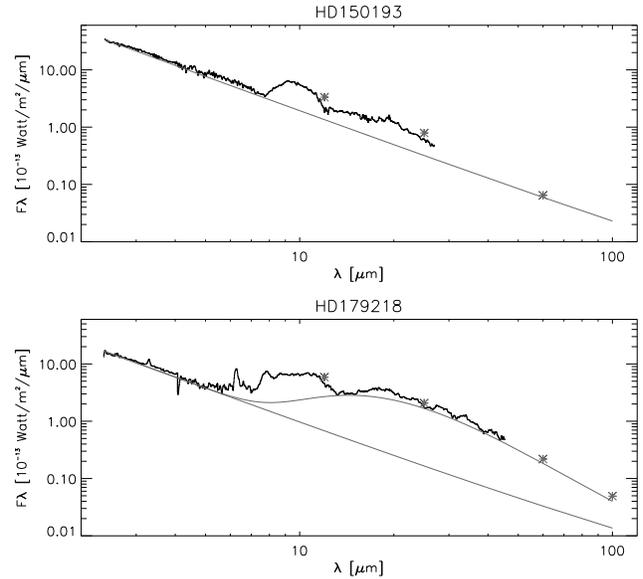}}}}
\caption{The determination of the continuum for the sources HD150193 (group 
IIa) and HD179218 (group Ia) in log $\mathrm{F_{\lambda}}$ versus log $
\lambda$ space. Full line: SWS spectrum, straight dotted line: power-law 
continuum fit; curved dotted line: sum of a power-law and a BB (T $\simeq$ 
190 K) continuum; asterisks: IRAS colour-corrected fluxes.}
\label{loglog}
\end{figure}

Notwithstanding these large differences, the overall structure of the
dust discs seems to be similar. It is possible to decompose the
spectra into at maximum three components: a power-law, a black body
(BB) and the solid state bands. As a first step, we fitted the IR
continuum of the stars showing a flat continuum with a power-law.
Actually, the determination of the continuum is non-trivial and should
be taken with some caution.  After some experiments, we found that the
continuum can be best determined by plotting the spectrum as log
$\mathrm{F_{\lambda}}$ versus log $\lambda$. As an example, we show in
Fig.~\ref{loglog}, upper panel, how the continuum determination was
done for HD150193. It is surprising to see that for at least six
sources (HD104237, HD142666, HD144432, HD150193, HD163296 and 51 Oph),
the continuum can be fitted very well with a power-law. We have to
remark here that we did not remove the photospheric component, since
it is only a negligible ($<$ 10 \%) fraction of the total flux. Only
for one source, 51 Oph, the dust component is less dominant, and the 
photospheric contribution to the total flux in the IR is much more
important. We therefore did not determine the power-law continuum for
this source.

\begin{figure}
\resizebox{\hsize}{!}{{\includegraphics{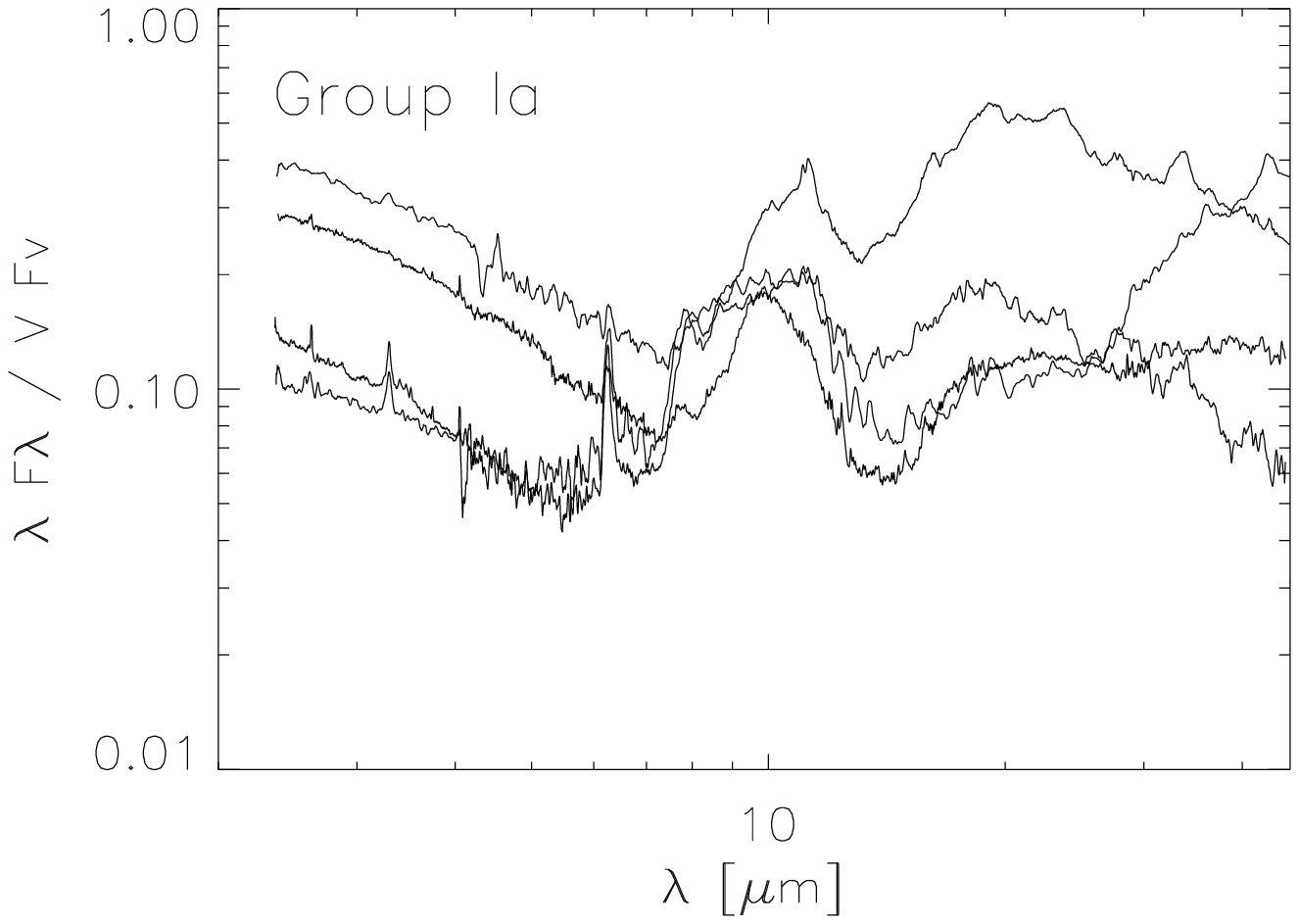}}}
\resizebox{\hsize}{!}{{\includegraphics{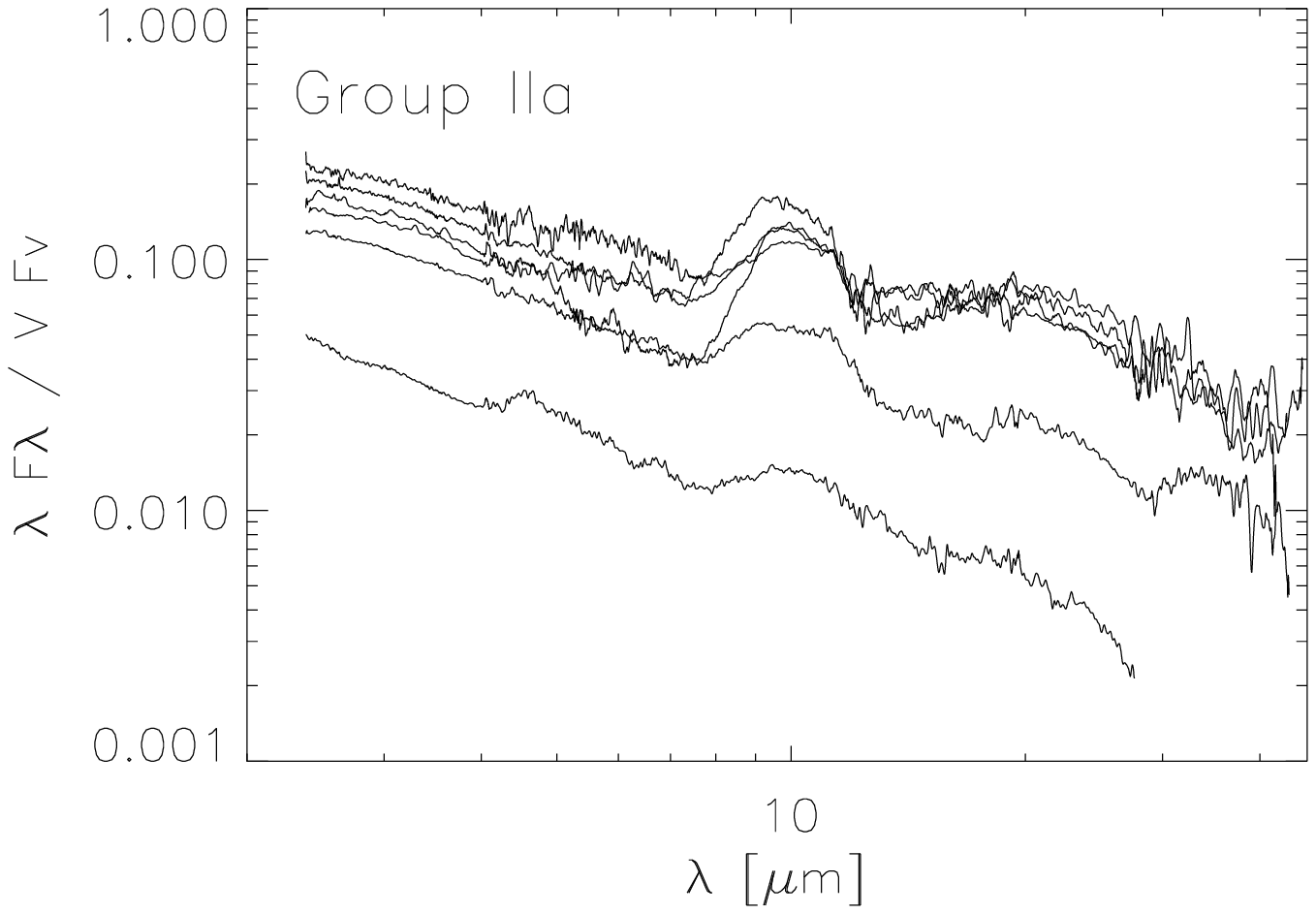}}}
\resizebox{\hsize}{!}{{\includegraphics{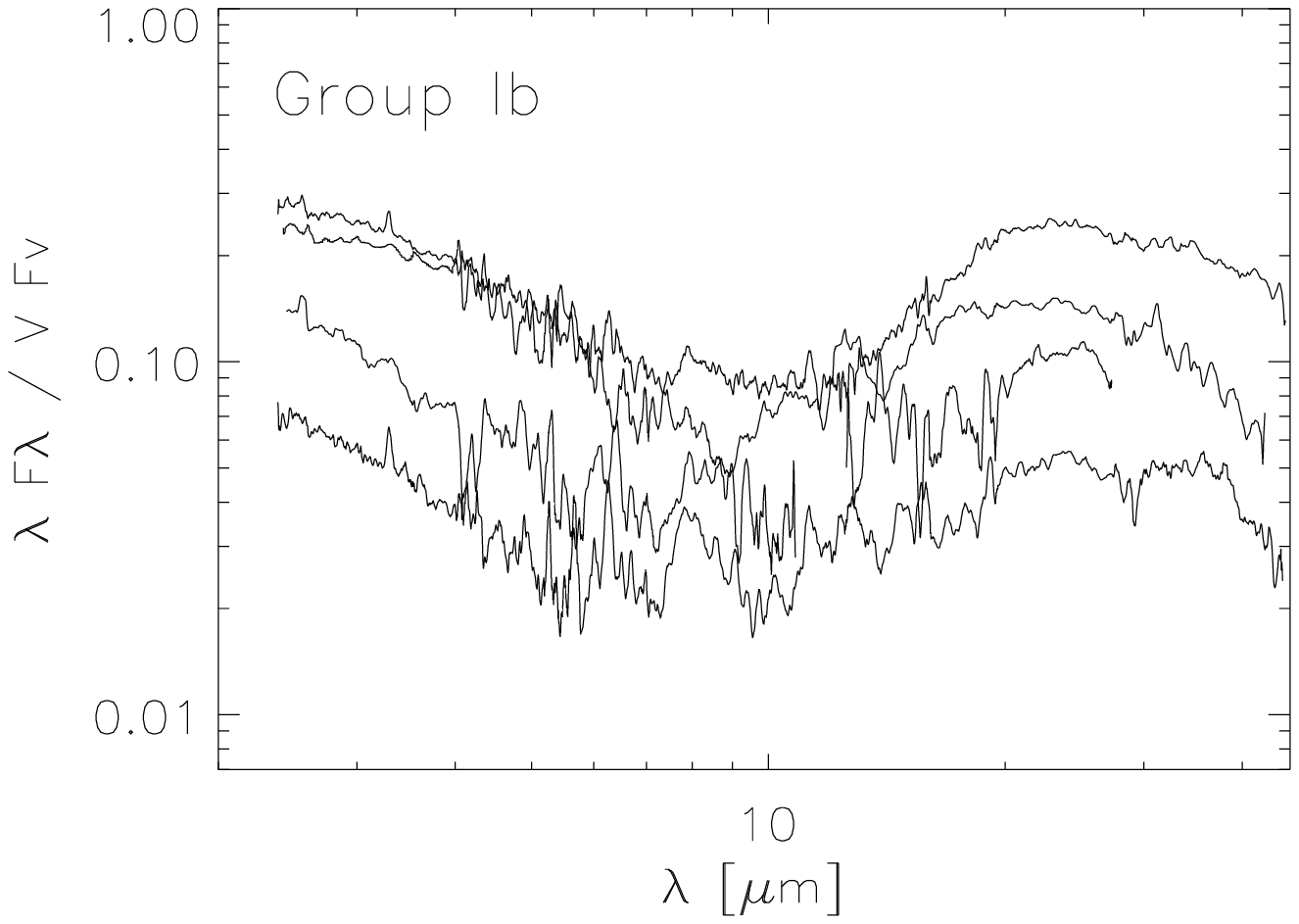}}}
\caption{\label{sws1}ISO spectra of the fourteen sample HAEBE stars, ordered 
by group. Group Ia: AB Aur, HD100546, HD142527, HD179218; group Ib: HD100453, 
HD135344, HD139614 and HD169142; and group IIa: HD104237, HD142666, HD144432, 
HD150193, HD163296 and 51 Oph. }
\label{3isoplotjes}
\end{figure}

\begin{figure}
\resizebox{\hsize}{!}{{\rotatebox{90}{\includegraphics{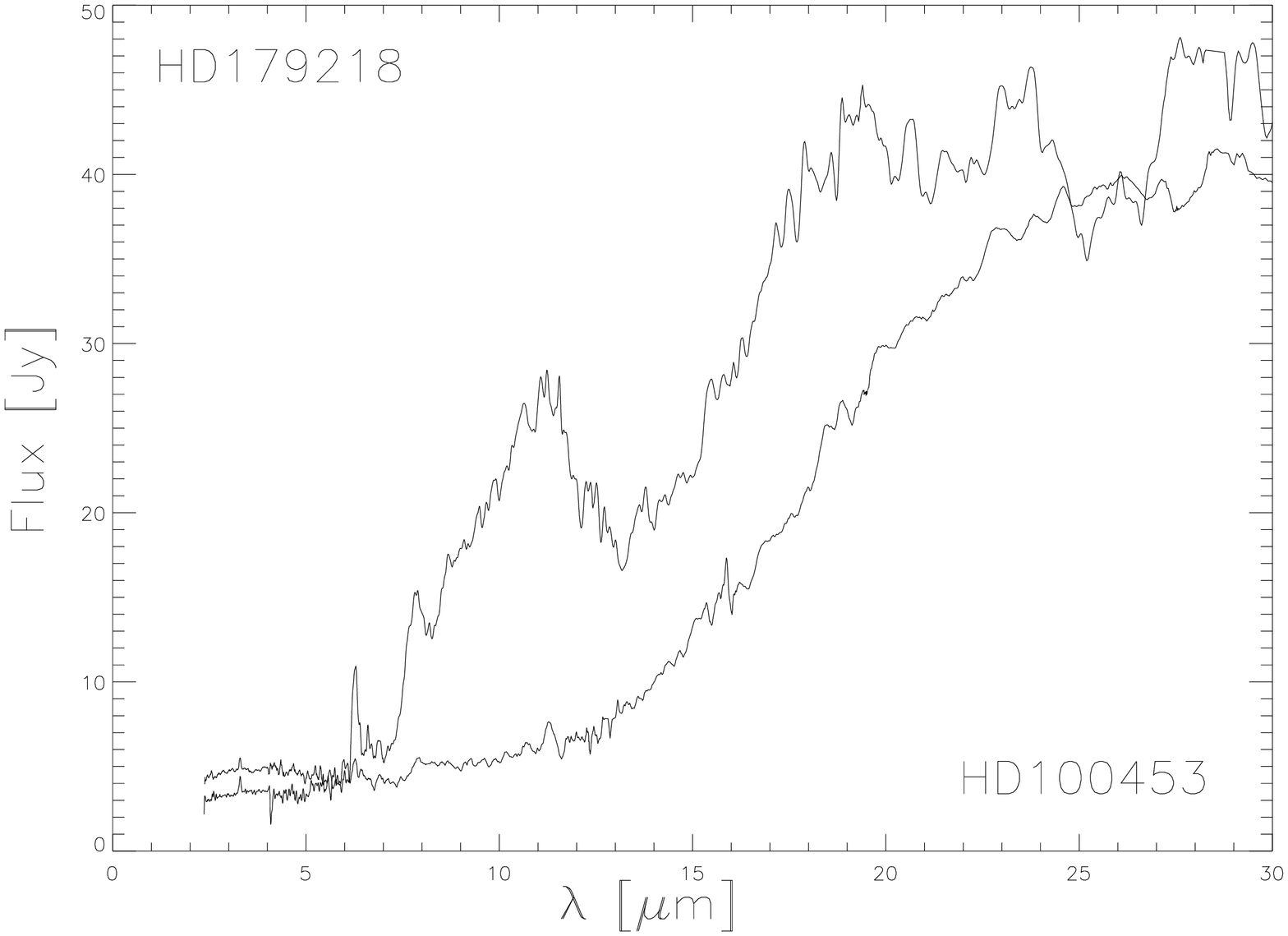}}}}
\caption{The spectrum of HD179218 compared to that of HD100453. The spectrum
of HD179218 can be obtained by adding silicate emission bands to the spectrum 
of HD100453.}
\label{versus}
\end{figure}  

We then proceeded to fit the sources with a rising mid-IR
continuum, assuming we could apply a similar power-law continuum fit
for these sources. This assumption is supported by the similarity in
the near-IR region for all our sample stars. For these sources, such
as HD179218 (see Fig.~\ref{loglog}, lower panel), an additional black
body\footnote{This BB component is usually a single-temperature BB, 
but it also can be a set of typically 2-3 BBs with different temperatures, 
to which we refer to as 'the' BB component for convenience, as an analogue 
to the power-law continuum.} (BB) on top of a power-law is needed to fit the 
continuum. Remarkably, with just these two components (a power-law
and a BB), the continuum of all sources can be fitted. The slope of
the power-law fits are listed in Table~\ref{submm}, together with the
average temperature of the respective BBs, when needed. In the sub-mm
region, we observe a turn-down in the slope of the continuum. This is 
because sub-mm wavelengths are longwards of the peak of the black body
for the coolest grains in the disc (thus we observe a Rayleigh-Jeans tail 
in the SED).

We have accordingly classified the sample: group I contains sources
for which the continuum can be reconstructed by a power-law and a black body,
and group II sources only need a power-law to fit their continuum. The
groups can be further subdivided according to the presence or absence
of solid state bands. In group Ia/IIa sources solid state bands are
present, while group Ib/IIb sources are without solid state bands. Our
sample does not contain a source which would fit in group IIb,
i.e. there is no star in our sample with a pure power-law continuum
which lacks solid state bands. This can be an observational selection
effect, as sources without a BB continuum are already fainter than
others. We thus have three distinct groups in our sample. In
Fig.~\ref{3isoplotjes} we display the combined SWS spectra of the
fourteen HAEBE objects arranged in these three groups.

If we neglect the silicates, then the shape of the IR spectra of group
Ia and Ib sources is very alike. They both have a prominent cool dust
component, rising in the mid-IR. What remains of the group Ib spectra
after continuum subtraction is essentially the same as for those of
group Ia sources without their silicate features. This is shown in
Fig.~\ref{versus}, where we compare HD179218 (group Ia) with HD100453
(group Ib). Adding the right amount of amorphous and crystalline
silicate components, we can obtain the spectrum of HD179218 starting
from HD100453. It is striking that both groups seem to have a similar
continuum, yet the silicates behave completely different in both sources.

To {\em summarize} these observational data, it first appears that
both the near-IR and sub-mm excesses are similar for all stars in our
sample; the mid-IR flux, on the contrary, shows large source to source
differences. Furthermore, the spectra can be decomposed into at maximum
three parts: a power-law, a BB and the solid state bands.

\begin{center}
\begin{tiny}
\begin{table}
\caption{Spectral slope of the power-law continuum fits in the IR region 
and the slope of the continuum in the sub-mm region. The slopes were 
measured in a log $\lambda F_{\lambda}$ versus log $\lambda$ scale. For
51 Oph, we could not determine a power-law continuum in the IR. The 
temperature of the BBs needed to fit the remains of the power-law subtracted 
continuum are listed as well. HD100453 and 51 Oph have no sub-mm slope 
listed because there are no 3 $\sigma$ sub-mm measurements available for 
these stars. For stars indicated with a '*', we had to include an IRAS 
flux (100~\mic~flux for HD100546 and HD104237; 60~\mic~flux for HD150193)
in the determination of the sub-mm slope, as there is only one sub-mm 
measurement available for those stars. }
\label{submm}
\begin{tabular}{clrrr}
\hline
\hline  \noalign{\smallskip}
Group&Source &IR   &BB       &sub-mm\\
     &       &slope&Temp. (K)&slope\\
\hline  
\hline \noalign{\smallskip}

Ia &AB Aur  &-1.20& 93   &-4.28 \\
Ia &HD100546&-1.17&170   &*-3.08\\
Ia &HD142527&-1.15& 73   &-3.60 \\
Ia &HD179218&-1.09&195   &-4.28 \\
\hline
\rule[3mm]{0mm}{1mm}
Ib &HD100453&-1.03&148   &-     \\
Ib &HD135344&-1.43&  -   &-3.28 \\
Ib &HD139614&-1.31&169   &-3.20 \\
Ib &HD169142&-1.25&  -   &-3.24 \\
\hline
\rule[3mm]{0mm}{1mm}
IIa&HD104237&-1.10&absent&*-2.81\\
IIa&HD142666&-1.03&absent&-2.91 \\
IIa&HD144432&-1.32&absent&-2.65 \\
IIa&HD150193&-0.95&absent&*-2.67\\
IIa&HD163296&-1.07&absent&-2.94 \\
IIa&51 Oph  &-    &absent&-     \\[1ex]
\hline
\end{tabular} 
\end{table}
\end{tiny}
\end{center}
%average of the ir slopes is 1.16 <-> 1.33 voor passive opt. thick disc

\subsection{Properties of the cold grains: mass and grain size}

The sub-millimetre flux is substantial in all our programme stars,
with the exception perhaps of 51 Oph. This implies that a large 
amount of cold, large grains must be
present further away from the star. The group-averaged sub-mm excesses
are comparable. To compare the mass of the cold dust between different
groups, we calculated $\mathrm{F_{800 \mu m}D^{2}}$, a normalization
of the cold dust mass which assumes that the stars are at the same
distance. These data are also listed in Table~\ref{sterren}.  The
amount of cold dust does not differ substantially between the three
groups.

A first indication of the size of the cold grains can be obtained by
inspecting the far-IR to sub-mm region. All our programme stars show a
turnover in their SED at far-IR wavelengths, indicating that at sub-mm
wavelengths we observe the Rayleigh-Jeans (R-J) SED of the ensemble of
cold grains present in the disc. If these grains are large compared to
the wavelength at which they radiate, the spectral slope will be
$\lambda F_{\lambda} \propto \lambda^{-3}$, while for small grains
this slope will be equal to $\lambda F_{\lambda} \propto
\lambda^{-(3+p)}$, where p is defined from $Q_{\nu} \propto \nu^{p}$,
it is the slope of the emissivity law of the dust at sub-mm
wavelengths. Estimating the size based on spectral indices should be
done with some caution, since the grain emissivity and the temperature
distribution of the grains in the disc also affect this slope. 

Spectral sub-mm slopes have been determined for our sample and are
shown in Table~\ref{submm}. Our sample shows a fair similarity in
spectral slope, except for the stars AB Aur and HD179218. For AB Aur,
the determination of the slope is more accurate than for HD179218,
because the last source has only two sub-mm points at a small
$\lambda$-interval. As was already noted by van den Ancker et
al. (1999), AB Aur has a very steep spectral slope ($\lambda
F_{\lambda} \propto \lambda^{-4.3}$); both stars jump out when
compared to the average of our sample ($\propto \lambda^{-3.2}$); the
slope of AB Aur and HD179218 is significantly steeper than that of the
tail of a black body ($\propto \lambda^{-3.0}$). Bouwman et al. (2000)
show that in AB Aur the 10~$\mu$m~silicate band and the mid-IR
continuum are dominated by micron-sized grains, while the millimetre
continuum is produced by grains with typical sizes of several 100
microns. These grains are optically thin at sub-millimetre
wavelengths, contrary to millimetre-sized grains around other HAEBE
stars (such as e.g. HD163296, van den Ancker et al. 2000). Possibly
the grain size distribution in HD179218 is more similar to that of AB
Aur, with somewhat larger grains (but still in the range of $\sim$
100~$\mu$m~size) producing the sub-millimetre flux. Sylvester et
al. (1996) already noted from their mm/sub-mm photometry that the dust
grains around most of their Vega-like systems are much larger than
those found in the interstellar medium (sub-\mic~sized, Mathis et al. 1977).

%%%%%%%%%%%%%%%%%%%%%%%%%%%%%%%%%%%%%%%%%%%%%%%%%%%
\section{ISO-SWS spectra and their wealth of features}

\begin{table*}
\caption{Presence of the solid state and PAH bands. \su: detection, 
?: possible detection, -: no detection. Regions for which we have no data or 
which we can not trust because of poor quality are indicated with n.a. 
(not available).}
\label{pahs}
\scriptsize
\begin{tabular}{lccccccccccccccc}
\hline
\hline \noalign{\smallskip}
Feature          &PAH  &PAH  &PAH      &PAH  &Si-O&PAH${^a}$&Oliv. &Oliv. &O-Si-O&Oliv. &FeO &Oliv.&Oliv. &Oliv. &$\mathrm{H_{2}O}$ ice \\
$\lambda~[\mu m]$&[3.3]&[6.2]&[``7.7'']&[8.6]&[9.7]&[11.2]  &[11.3]&[16.3]&[19]  &[19.8]&[23]&[23.8]&[27.9]&[33.7]&[43.8]\\[1ex]
\hline
\hline \noalign{\smallskip}
 AB Aur          &-    &\su  &\su      &\su  &\su  &\su     &-     &-     &\su   &-     &\su &-      &-     &-     &-  \\   
 HD100546        &\su  &\su  &\su      &\su  &\su  &\su     &\su   &\su   &\su   &\su   &\su &\su    &\su   &\su   &\su\\   
 HD142527        &\su  &\su  &-        &-    &\su  &?       &\su${^b}$&-  &\su   &-     &\su&-     &?     &?     &\su${^c}$\\   
 HD179218        &\su  &\su  &\su      &\su  &\su  &?       &\su${^b}$&?  &\su   &\su   &?   &\su    & ?    &\su   &\su\\[1ex]
\hline \noalign{\smallskip}
 HD100453        &\su  &\su  &?        &-    &-    &\su     &-     & -    &-     &-     &-   &-      &-     &-     &-  \\ 
 HD135344        &?    &-    &?        &-    &-    &?       &-     & -    &-     &-     &-   &n.a.   &n.a.  &n.a.  &n.a.\\ 
 HD139614        &-    &-    &-        &-    &-    &-       &-     & -    &-     &-     &-   &-      &-     &-     &-\\
 HD169142        &\su  &\su  &\su      &?    &-    &\su     &-     & -    &-     &-     &-   &-      &-     &-     &-\\[1ex]
\hline \noalign{\smallskip}
 HD104237        &-    &-    &-        &-    &\su  &-       &?     &-     &\su   &\su   &?   &?      &?     &?     &-\\ 
 HD142666        &\su  &?    &\su      &?    &\su  &\su     &\su${^b}$&-  &\su   &-     &\su &?      &-     &?     &-\\
 HD144432        &-    &-    &-        &-    &\su  &-       &?     &-     &\su   &-     &\su &n.a.   &-     &n.a.  &n.a.\\ 
 HD150193        &-    &-    &-        &-    &\su  &-       &\su   &-     &\su   &-     &?   &n.a.   &n.a.  &n.a.  &n.a. \\ 
 HD163296        &-    &-    &-        &-    &\su  &-       &\su   &\su   &\su   &\su   &\su &?      &?     &?     &? \\  
 51 Oph          &-    &-    &-        &-    &\su  &-       &?     &-     &\su   &-     &?   &n.a.   &n.a.  &n.a.  &n.a.\\[1ex] 
\hline
\end{tabular} 
\footnotesize{
\\${^a}$  possible blend with the olivine 11.3~\mic~band\\
 ${^b}$ possible blend with the PAH 11.2~\mic~band\\
% ${^c}$ band seems to be shifted to 23~\mic. \\
 ${^c}$ doublepeaked at 43.8~\mic, probably also due to montmorillonite 
(Malfait et al. 1999a).}
\end{table*}

\begin{figure*}
\resizebox{\hsize}{!}{{\rotatebox{90}{\includegraphics{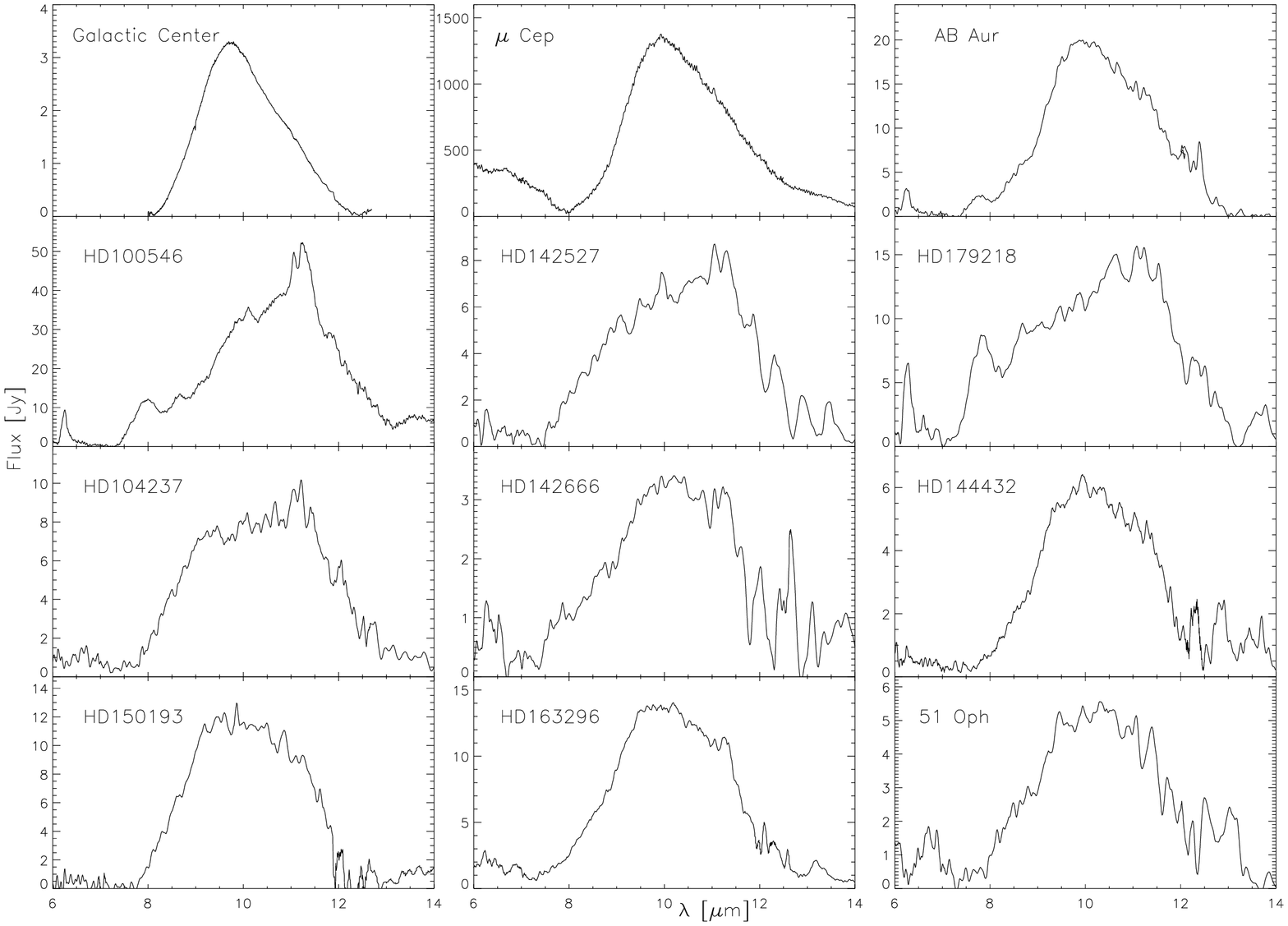}}}}
\caption{Continuum subtracted spectra of the 10 \mic~silicate feature. We 
also show the spectra of the M supergiant $\mu$ Cep and the galactic center 
for comparison. The galactic center spectrum has been converted from 
absorption to optical depth, which is plotted on the y-axis. Amorphous 
silicate, as seen in the Galactic Center spectrum, peaks at 9.7~\mic, 
while crystalline silicates peak at longer wavelengths ($\sim$ 11~\mic). 
Notice the quasi omnipresence of an 11.3~\mic~band, due to crystalline 
silicates and/or PAHs in the spectra of the Herbig Ae/Be stars. Only AB 
Aur has no crystalline signatures at all. Group Ib sources are not shown due 
to a lack of band structure in this region.}
\label{silirest}
\end{figure*}

\begin{figure*}
\resizebox{\hsize}{!}{{\rotatebox{90}{\includegraphics{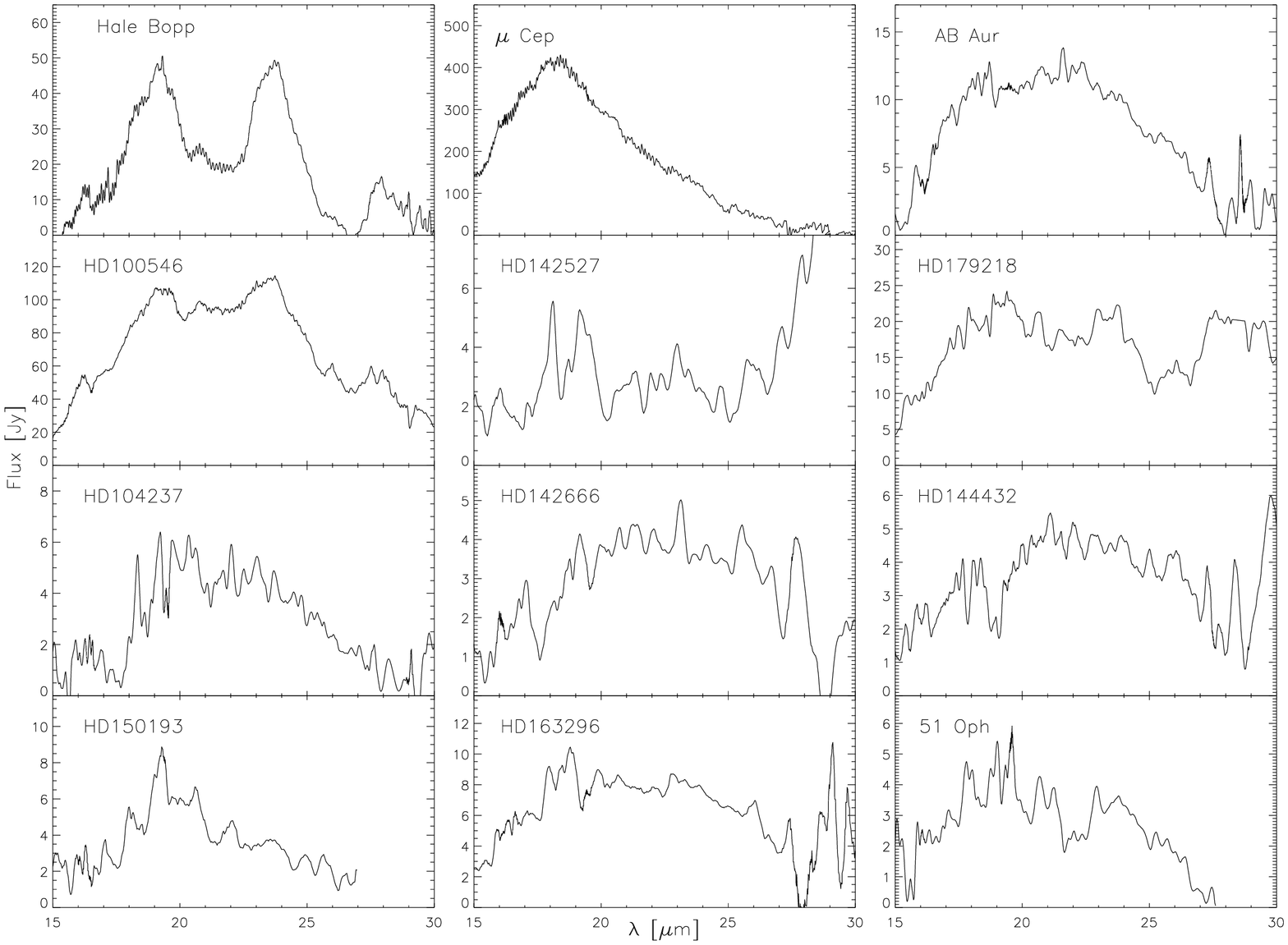}}}}
\caption{Continuum subtracted spectra of the 15 to 30 \mic~region.
We also show the spectra of the M supergiant $\mu$ Cep (amorphous silicates)
and comet Hale-Bopp (crystalline silicates) for comparison. From a 
comparison with $\mu$ Cep, it is clear that most sources need an additional
band longwards of the 18~\mic silicate feature to explain their spectra. 
Group Ib sources are not shown due to a lack of band structure in this region. }
\label{siliwarm}
\end{figure*}

We describe general trends in the appearance of the solid state bands
we have detected in our sample of stars. We have carefully inspected
the individual spectra in order to verify the reality of solid state
bands. This can be done by analyzing the individual detector scans,
and by inspecting the two independent scan directions at which the
data were taken. We list the solid state bands and their possible
identifications in Table~\ref{pahs}. The features for which we have an 
unsure identification are listed with a question mark. In particular, 
the PAH feature at 11.2~\mic~can be easily confused with the olivine 
feature at 11.3~\mic. Caution should also be taken around 12~\mic, 
where an SWS band jump occurs. In Figs. \ref{silirest} and
\ref{siliwarm}, we show the continuum subtracted spectra in the 6 to
14~\mic~wavelength region and the 15 to 30~\mic~wavelength region
respectively, highlighting the solid state bands found in these
spectral regions. Below, we briefly discuss the different solid state
components.

\subsection{The 2-7 micron wavelength region}

From Fig.~\ref{3isoplotjes}, one can observe immediately that there is 
a striking similarity between the spectra from 2 to 7~\mic~in 
all three groups, indicating a similar composition and temperature 
distribution of the material emitting in this range. This near-IR excess 
must be caused by small hot grains, most likely metallic Fe (or FeO), or 
a carbonaceous component, e.g. graphite or amorphous carbon (van den Ancker 
et al. 1999, Bouwman et al. 2000), as other materials would not 
survive the high temperatures ($\ga$ 1000 K) close to the star.
 
\subsection{The PAH bands}

PAH bands are present in group Ia and Ib sources, and only weakly in
one source of group II (HD142666); they are seen in at least 50\% of
our total sample. This number might still go up as weak features can
be lost in the noise. PAHs are strongest in early type sources, as 
can be expected (since they are excited by UV radiation).
That the PAHs most probably belong to the stellar
environment can be ascertained from e.g. ISOCAM data of HD 179218, a
source with is not extended (Siebenmorgen et al. 2000). Another
interesting observation is their different appearance, at e.g.
$\lambda \sim$ 6.2~\mic. PAHs in the ISM and in HII regions peak at
$\lambda$ $\sim$ 6.20~\mic, while at our stars they rather peak at
$\lambda$ $\sim$ 6.25~\mic. A preliminary result in this context is
that non-extended sources tend to have PAHs at 6.25~\mic~(Van
Kerckhoven, private communication). A more detailed study will follow
(Van Kerckhoven C., in preparation; Peeters E., in preparation).

\subsection{The silicates}

In Fig.~\ref{silirest} and \ref{siliwarm}, we show the regions
surrounding the strongest amorphous silicate bands. We also included a
spectrum of the M supergiant $\mu$ Cep (Kemper, private communication)
and the galactic center (Lutz et al. 1996, Tielens, private
communication), as prototypes of amorphous silicate. The spectra of
group Ib sources are not shown, as they do not show silicate
bands. The 8-12~\mic~silicate features are significantly different
from the so-called 'astronomical silicate', in the sense that they
peak at 11~\mic \ rather than at 9.7~\mic.  The only exception is AB
Aur, a group Ia source. The peak shift is attributed to a change in
the grain size distribution towards larger (\mic-sized) grains, and/or
the presence of crystalline olivine, causing a peak at
11.3~\mic~(Bouwman et al., in preparation). 

In some sources such as
HD100546, there is a large amount of crystalline silicates (Malfait et
al. 1998a), while in other sources, such as HD163296, it is less
so. That both AB Aur and HD100546 are members of group Ia, shows that
the crystallization degree of the silicate material is independent of
the shape of the overall spectrum. The interpretation from sub-mm data
that AB Aur has the least processed dust (smallest grains at sub-mm
wavelengths) is here further supported by the lack of crystalline
features in its spectrum. We, however, derived a similar slope for the
sub-mm region of HD179218 (see Sect. 2.2), a source with a large 
amount of crystalline silicates. The determination of this slope was less 
accurate, but it is for sure steeper than the other sources. From this we 
can infer that coagulation and crystallization processes occur on different
time-scales.

\subsection{The 23 micron feature}

The [15-30]~\mic~spectra (see Fig.~\ref{siliwarm}) of group Ia and IIa
sources are even more diverse than the [6-14]~\mic~spectra. From the
spectrum of $\mu$ Cep, we can see that the amorphous silicates alone
can not account for the broad bands seen in other sources, e.g. AB Aur
and HD142666. An additional component, around 22-28~\mic~must be
present as well. This component can be attributed to FeO (van den
Ancker et al. 2000) and/or crystalline silicates. 51 Oph shows two
clearly separated bands, supporting our interpretation of the two
component broad band. The same region for group Ib sources can be
fitted very well with black bodies with a temperature ranging between
150 and 170 K. Also here we see no indication for silicate bands in
group Ib sources.

\subsection{Notes on individual sources}

{\bf Group Ia}:\\
\smallskip\\
{\bf AB Aur}: The 10~\mic~region of AB Aur shows a broad silicate
feature superposed on a rising continuum. The feature peaks at
9.7~\mic, which is typical for amorphous silicate. It is the only
source in our sample to show merely pure amorphous silicates. The
20~\mic~region shows a broad band, probably due to a combination of
silicates and iron oxide. PAH bands are moderately strong, and present
at 6.2, 7.7, 8.6 and 11.2~\mic. This star has the steepest spectral
slope at sub-mm wavelengths. For a thorough analysis of this star, we
refer to van den Ancker et al. (1999) and Bouwman et al. (2000). \\
{\bf HD100546}: This star is thoroughly analyzed by Malfait et al. 1998a.
Amorphous silicate emission bands are visible at 10 and 18~\mic. It has the
largest amount of crystallinity in our sample, and shows strong PAH
bands at 3.3, 6.2, 7.7, 8.6 and 11.2~\mic. The similarity with the SWS 
spectrum of comet Hale-Bopp is remarkable (Waelkens et al. 1998). The 
comet Hale-Bopp is an end result of grain processing in our solar system.
Given the large similarities in spectral appearence with HD100546, we
can assume that the same kind of dust processing has taken place in 
both sources.\\ 
{\bf HD142527}: The silicate feature at 10~\mic~is shifted to longer 
wavelengths, pointing to the presence of crystalline silicates. The 
continuum rises from longer wavelengths on than most sources, pointing 
to colder dust. Weak PAH bands are present at 3.3, 6.2 and 11.2~\mic.  
We refer to Malfait et al. (1998a) for more details. This star is special
because it has a large $\mathrm{L_{IR}/L_{*}}$ ratio, typical for stars
with an active disc. But these stars also show evidence for outflow,
what is not observed in HD142527. Another explanation for the large IR 
luminosity is the presence of a more embedded companion, but this still 
needs to be investigated.\\
{\bf HD179218}: The overall shape of this spectrum is quite similar to 
that of HD100546. It shows the 10~\mic~silicate feature, and has a rising 
continuum. The abundance of crystalline silicates is less, but still quite 
high. This source has a higher abundance of crystalline pyroxenes than 
crystalline olivines, unlike observed in other sources (Malfait 1999b). 
Strong PAH bands are present at 3.3, 6.2, 7.7, 8.7 and 11.2~\mic.\\
\smallskip\\

{\bf Group Ib}:\\
\smallskip\\
{\bf HD100453}: This star does not show silicate emission bands.
It has a rising continuum and moderately strong PAH bands are present 
at 3.3, 6.2, 7.7 and 11.2~\mic.\\
{\bf HD135344}: In this source the silicates are absent. The spectrum
shows a rising continuum, but beyond 27~\mic~the spectrum is no longer 
usable due to low S/N. Weak PAH bands are present at 3.3, 7.7 and 11.2~\mic.\\
{\bf HD139614}: Also here there are no silicate bands. The spectrum
shows a rising featureless continuum, PAH bands are absent.\\
{\bf HD169142}: Moderately strong PAH bands are present at 3.3, 6.2, 
7.7, 8.7 and 11.2~\mic.  The spectrum shows a rising spectrum without 
silicate emission bands.\\
\smallskip\\

{\bf Group IIa}:\\
\smallskip\\
{\bf HD104237}: The 10~\mic~spectrum shows a very strong silicate
feature, superimposed on a flat continuum. The feature peaks longwards
of 10~\mic, so that crystalline silicates and/or larger silicate grains 
must be present. PAH bands are absent.\\
{\bf HD142666}: The spectrum shows a flat continuum, upon which the
10~\mic~feature is superimposed, peaking at 10.3~\mic. Very weak PAH
bands are present at 3.3, 6.2, 7.7, 8.6 and 11.2~\mic. The
20~\mic~region shows a broad band, probably consisting of silicate and
FeO.\\
{\bf HD144432}: The spectrum of HD144432 is very similar to that of
HD142666, but PAH bands are absent. Interestingly, these 2 sources are
in general quite similar, the most important difference being the
inclination of their disc (Meeus et al. 1998). The 20~\mic~region shows 
a broad band, probably consisting of silicate and FeO.\\
{\bf HD150193}: This source has a very strong 10~\mic~feature,
superimposed upon a flat continuum. The shape of the spectrum around
20~\mic~is very peculiar, and we are not sure if it is an artefact
or real. PAH bands are absent.\\
{\bf HD163296}: The 10~\mic~silicate feature is superposed on a flat
continuum. The 20~\mic~region shows a broad band, probably consisting
of silicate and FeO. PAH bands might be present, but we cannot
ascertain this. We refer to van den Ancker et al.  (1999) and Bouwman
et al. (2000) for more details and modelling.\\
{\bf 51 Oph}: This source is the most extreme, in the sense that its
turn-over point towards the R-J tail already starts at $\sim$ 30~\mic;
from which we can conclude is has a smaller amount of dust.
It also has a very rich spectrum, with both gas ($\mathrm{CO_{2}}$ at
4.2~\mic~and $\mathrm{H_{2}O}$ at 5-6~\mic) and solid state bands.
The 10~\mic~silicate feature is strong and imposed on a
descending continuum. The 20-30~\mic~region shows two clearly
separated bands, which we can attribute to amorphous silicate
(18~\mic), and to FeO or crystalline silicate (23~\mic). The
evolutionary status of 51 Oph is not very clear.  Some authors
classified this object as a Be star (Slettebak 1982), while others
consider it as a Herbig Ae/Be star (Malfait et al. 1998b) or even a Vega-type
star (Sylvester et al. 1996). A more detailed study of this object will be 
presented elsewhere (Meeus et al./van den Ancker et al., in preparation).

%%%%%%%%%%%%%%%%%%%%%%%%%%%%%%%%%%%%%%%%%%%%%%%%%%%
\section{Discussion}

Two important results of the comparative study of the SEDs of the
programme stars concern the near-IR and the sub-mm excess: 1. a
similar near-IR (1-8~\mic) excess is observed for all stars. Small hot
grains close to the star must be replenished because they are
continuously destroyed by the UV radiation of the star. The overall IR
spectrum being so diverse, but the near-IR so similar indicates that
the material close to the star is homogenized; 2. the sub-mm excess is
substantial for all stars in our sample. This implies that large
grains already formed when the star formation process comes to an end,
and that large grains remain present around the stars during their
evolution towards Vega-type stars. In a survey of T Tauri Stars (TTS),
Beckwith et al. (1990) also found that the disc mass does not decrease
with increasing stellar age. Warm grains, however, seem to disappear
on a shorter time-scale.

The third remarkable observation is the strong variation in strength
of the silicate feature: for some stars it dominates the ISO spectrum,
for others it is moderately strong, and for some stars it is even
absent. It is surprising that there is no relation between the
silicate feature at 9.7~\mic~and the near-IR excess, although both
emission features must be caused by hot material, presumably located
in the same region, and consisting of grains of a similar size.  We
can already exclude inclination angle effects since our sample
includes two objects, HD142666 and HD144432, with a very different
inclination (Meeus et al. 1998) but with almost identical ISO spectra,
both showing a prominent 10~\mic~silicate band. There are several
possible explanations to our observations, and these 'scenarios' will
need to be confirmed by more detailed observations and by careful
modelling of the CS material. In the following subsection we propose a
global model.

\subsection{Geometry of the disc and its effects on the SED}

As stated before, the main difference between group I and II sources
is the amount of mid-IR excess, which is dominant and rising for group
I sources, while moderate and rather descending for group II sources.
Both groups also differ as far as the total IR luminosity is
concerned: $\mathrm{L_{IR}/L_{*}}$ is on average 0.52 and 0.17 for
group I and II respectively (see Table~\ref{sterren}). Group II
sources thus have the smallest emitting surface, and probably the
smallest mass of warm dust. On the other hand, the solid state bands
are present with equal average strength in group Ia and IIa sources:
the solid state bands thus must be formed in yet another region. Natta
et al. (2000) calculated the silicate 10~\mic~feature intensity with
the models of Chiang \& Goldreich (1997), and they needed to add a
power-law component to the emission of the disc atmosphere to fit
their 10 micron spectra of T Tauri Stars; this finding supports our
distinction between the region in which the solid state bands are
formed and the region from which the power-law emission originates.

\begin{figure}
 \resizebox{\hsize}{!}{{\rotatebox{0}{\includegraphics{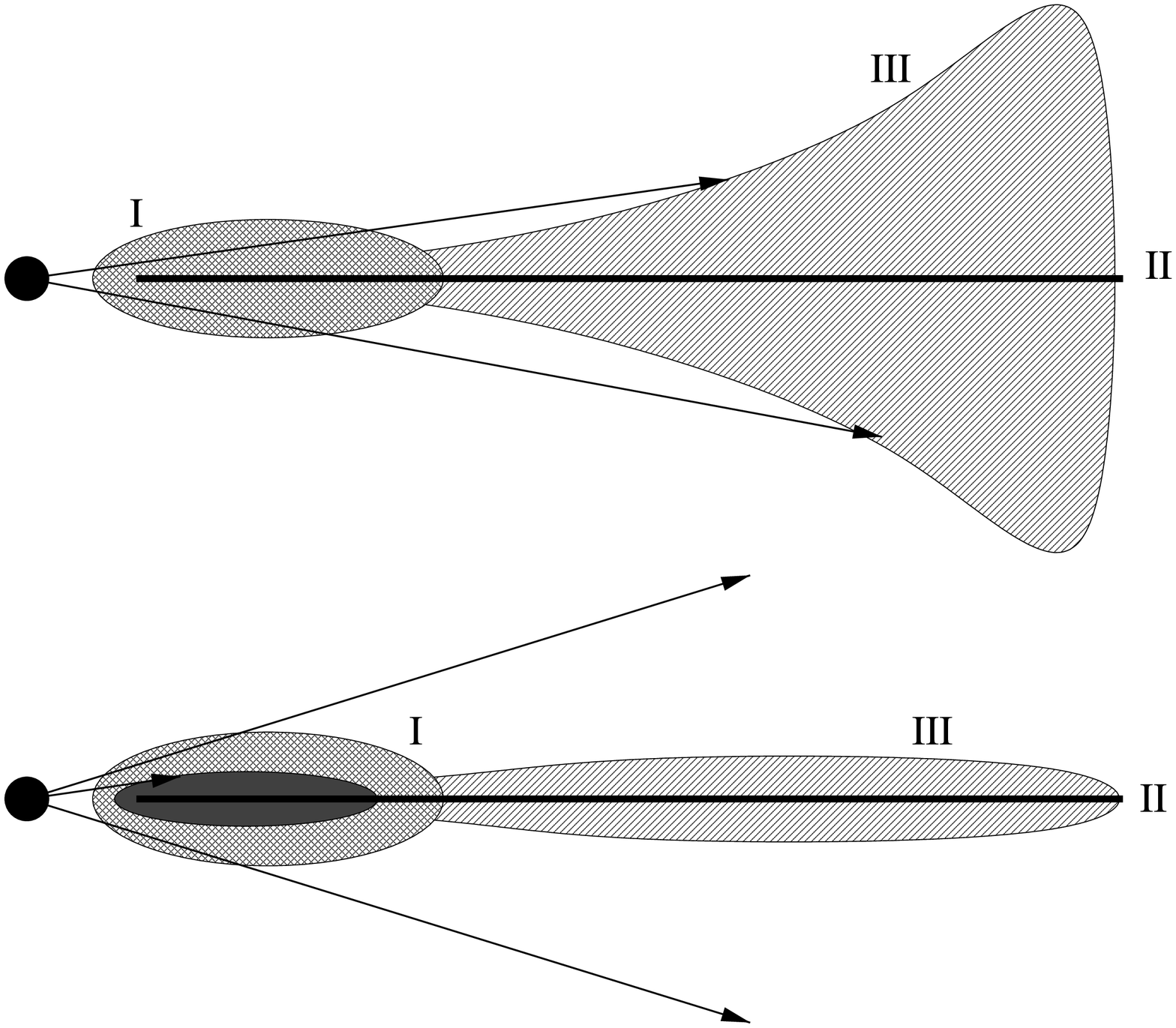}}}}
\caption{Schematic presentation of the model. The disc consists out of three
parts: I) a (partially) optically thin inner part ($\sim$ 10 AU); II) a 
geometrically thin, optically thick midplane; and III) a flaring part. The 
upper panel shows an entirely optically thin inner disc, where the star 
illuminates the disc's surface and causes the flaring. The lower panel shows 
the case where the surface of the disc is shielded from direct stellar 
radiation by an optically thick region, so that it does not flare (Bell et 
al. 1997, Nelson et al. 2000).}
\label{model}
\end{figure}  

The classification of our sample of HAEBE stars into two main groups
could be explained with the following simple physical picture, which
is shown schematically in Fig.~\ref{model} (upper panel: a group I
source, lower panel: a group II source) and consists of the following
components: I) a (partially) optically thin inner part ($\sim$ 10 AU
in size); II) an optically thick, geometrically thin disc which forms
in an early stage; and III) a flaring part, exposed to stellar
radiation.  Similar geometries were suggested in studies of
protostellar accretion discs by e.g. Bell et al. (1997) and Nelson et
al. (2000). There it is shown that the protostellar discs can have an
inner disc ($\sim$ 10 AU in size) with increased scale height
(i.e. "puffed up") which can shield the outer parts of the disc so
that no flaring occurs. One can explain the SEDs of group II sources
by assuming that the inner part (component I lower panel
Fig.~\ref{model}) is partially optically thick, shielding the outer
parts of the disc from direct stellar radiation, which prevents the
disc from flaring (hence one does not observe a BB component). SEDs of
group I sources then can be explained by an additional flared
component outwards of an optically thin inner disc which does not
shield. The non-occurence of a group IIb source could be a selection
effect, since one would expect such sources to have the lowest flux
levels.

The emission of an optically thick, passive disc varies as $\lambda
F_{\lambda} \propto \lambda^{-1.3}$ (Friedjung 1985). The average IR
slope for the sources in our sample is -1.2, very close to the value
for the slope of optically thick discs. There is, however, some
dispersion in our slopes, but nevertheless the assumption of an
optically thick disc to explain the power-law component seems to hold
when confronted with theoretical models.

In what follows we discuss the separate components of our model and
how they appear in the SED:
\begin{itemize} 

%hier ook zeggen dat pahs are transiently heated, dus kunnen verder
%weg zitten, maar grotere silicaten niet.

\item the {\bf geometrically thin}, optically thick midplane consists 
of large grains, and is responsible for the IR power-law and the sub-mm
continuum emission. Here resides the bulk of the material (Bouwman et
al. 2000). The small hot grains in the inner part of the disc are
responsible for the near-IR excess. The similarity of the spectra in this 
wavelength range suggests a very similar chemical composition and 
temperature distribution.

\item hot grains in a (partially) optically thin {\bf "disc
atmosphere"} cause the 10 and 18~\mic~amorphous silicate emission
bands.  That the solid state bands are indeed likely not to originate
from the flared region around the disc is evidenced by: 1) our group
IIa observations (solid state bands are present, but the BB caused by
the flaring is absent); 2) the fact that silicate grains causing the
10~\mic~emission are too large to be transiently heated, so they have
to be located close to the star; and 3) observations of evolved
objects by Molster et al. (2000), who notice that disc sources have a
high abundance of crystalline silicates.  Where this optically thin
region is located exactly is not clear, but it certainly must be close
to the disc and to the star.

\item the {\bf geometrically thick}, optically thin, flared dust layer
below and above the midplane is well mixed with the gas and contains
small, warm grains that cause the 100-200 K black body component. Only 
a small amount of the total disc mass is located in this flared region. In
group II sources this region must be very small or even absent, since
we only see a power-law continuum in the IR. This conclusion is also
supported by theoretical models from Kenyon \& Hartmann (1987), which
give an upper value for $\mathrm{L_{IR}/L_{*}}$ for flat discs of
0.25, while flaring discs have larger IR luminosities. When comparing
the values we determined for $\mathrm{L_{IR}/L_{*}}$ (see
Table~\ref{sterren}) with their predictions, we see that group II
sources indeed have $\mathrm{L_{IR}/L_{*}}$ values arguing for a flat
disc. The scale-height of the flaring determines the amount of mid-IR
excess: the larger the flaring, the stronger the excess. The
temperature of the BB is determined by the distance of the onset of
the flaring to the central star: the further away the flaring starts,
the colder the BB. If we e.g. compare HD179218 with HD142527, then the
flaring must start further away for the latter object, as evidenced by
its BB temperature (see Table~\ref{submm}). That PAHs are only present
in group I sources, with the exception of HD142666 (group IIa, but
only very weak PAH features), suggests that PAHs are most likely
located in the flared region, where they are exposed to the stellar UV
radiation.  They are small enough to be transiently heated, without
needing to be located close to the star.

\end{itemize}

Other geometries interpreting the SEDs of HAEBEs have been proposed,
e.g. Miroshnichenko et al. (1999) propose an envelope in addition to a
disc to model the dust emission from HAEBEs. We note, however, that
the highly abundant crystalline silicates in HD100546 (that have a
temperature of 200~K or less, Malfait et al. 1998a) are most probably
formed in the disc (Molster et al. 1999), suggesting that the grains
at this temperature are not in a loosely bound envelope, but are
intimately connected to the disc.  We suggest that the 200~K black
body component in this star, and by analogy in other HAEBE stars in
our sample, is associated with the (flared) disc and not with an
envelope. Waelkens et al. (1994) and van den Ancker et al. (1997)
interpret the apparent broad dip around 10 micron in the SEDs of
HAEBEs as a physical gap in the radial distribution of the CS dust
(there is simply no dust at a certain distance to the star,
corresponding with the region where the 10 micron flux should come
from). To cause a physical gap, another body surrounding the central
star must be present as well. In the light of these different
possibilities, detailed spatial information is essential to
disentangle the location of the different spectral components.

%[hd163296 heeft wel nog wat flaring maar ook minst cryst materiaal,
%terwijl hd104237 vlak blijft, ook bij iras metingen en wel al crystallijn
%is. dus mischien verband flaring en crystallijniteit? kan niet echt
%met hd100546!!!]
 
\subsection{Evidence for grain growth}

Evidence for grain growth in the discs of HAEBEs has been found
by several authors. Radiative transfer modelling by e.g. Bouwman et 
al. (2000) shows that the dust grains around the Herbig Ae stars AB Aur and 
HD163296 are much larger than those of the ISM. Furtheron, Grady et al. (1996) 
observe accreting CS gas in HAEBEs and attribute this to large infalling 
objects. A population of large ($\sim$ 0.1-1mm) grains is needed to explain 
the observed sub-mm fluxes in our sample. These observations show that grains 
growth indeed takes place around HAEBE stars, and that it is an on-going 
process which we can observe indirectly by looking at a large sample of 
objects.

We now consider the cause of the difference in SED between group I and
II sources. If our interpretation of a flared region above and below
the disc causing the BB component is correct, its absence in group II
sources may imply that these small, warm grains have been removed due
to coagulation and/or to radiation pressure exerted on the grains by
the central star, so that the small warm grains have slowly
disappeared and their scale-height has diminished accordingly. The
absence of PAH bands in group II stars already supports the assumption
that small grains are removed in the extended region. If we expect
grains to grow during the star's evolution towards the MS, then the
excess as a whole should decrease. This is consistent with the amount
of IR luminosity we derive from the SEDs: stars from group II show a
smaller $\mathrm{L_{IR}/L_{*}}$ ratio than group I stars, so that
group II sources may have the most evolved dust grains. This simple
evolutionary assumption is supported by observations of TTSs by
Beckwith et al. (1990), where it is shown that older discs tend to be
colder and less luminous. The grain-growth assumption to explain the
differences between group I and II also holds in the sub-mm region:
from Table~\ref{submm} it is clear that group II sources have a less
steep sub-mm slope than group I sources (on average -2.8 versus -3.6),
which means that the latter have smaller grains radiating in this
region.

% This scenario could also be tested by checking
%*on the gas content of the discs, which may then be lower for the group 2
%*sources compared to groups 1 and 3.   

To conclude, the SEDs are consistent with a disc model in which the
differences between group I and group II can be explained by a
different extent of the warm flared layer above and below the
mid-plane. Both the $\mathrm{L_{IR}/L_{*}}$ ratio and the sub-mm slope
suggest that group II sources have larger grains than group I
sources. These observations are consistent with an evolution from
group I to group II sources. However, a larger sample and spatial
information are needed to prove such an hypothesis.

\subsection{Influence of the stellar age on dust properties}

The age of the sample stars was derived by van den Ancker et al. (1998)
using Hipparcos parallaxes, and is listed in Table~\ref{sterren}.
Unfortunately, we do not dispose of ages for group Ib stars. We are
aware of the fact that our sample is biased towards the more evolved
sources, since our sources are isolated. However, we can already reach
some conclusions. There is no clear trend between age of the central
star and amount of crystalline material: the star HD100546 (showing the 
largest amount of crystalline dust) is probably the most evolved one, 
while HD179218 (probably the youngest source) also shows a substantial 
amount of crystalline dust. AB Aur, on the other hand, is also already 
more evolved,
but shows only evidence for amorphous silicates. The range in age is
very similar for group Ia and group IIa stars; from a confrontation
between both groups we can conclude that the {\bf stellar} age has no (or
little) influence upon: 1) coagulation, since group IIa stars have a
less steep sub-mm slope than group Ia stars; 2) amount of material in the
flared region, as this is much less or even absent in group IIa stars;
and 3) presence of PAH bands, as they are merely absent in group IIa
stars. We thus conclude that the timescales on which star and disc evolve
are not strongly coupled for the sample of HAEBEs studied here.

\subsection{The amorphous silicate behaviour}

It is surprising that the silicate feature is absent in group Ib
sources, as the rest of the SED argues for similar disc properties as
for group Ia sources. It is not unreasonable to assume that grains grow
during the stellar evolution towards the MS. Therefore, the absence of
the silicate feature could be easily explained by the absence of small
grains (with average sizes less than a few \mic). However, the
presence of a near-IR excess points to the presence of small, hot
grains. Furthermore, almost all of the group Ib sources show PAHs in
their spectra, which are also caused by very small particles. It thus
seems that the small silicate grains evolve differently than
other small particles. The explanation for group Ib sources could be
either that the inner part does not exists (e.g. could be
geometrically thin, optically thick), or that there are no small 
($\la$ 50~\mic) silicate grains.

Remarkably, in an atmospheric abundance analysis by the authors and
another analysis by Dunkin et al. (1997), a silicon depletion around 3 out of
4 of the group Ib HAEBE objects was revealed, while sources from group
Ia and IIa were shown to have solar abundances. The photospheric
silicon depletion for group Ib stars may further support that around
these stars, silicates behave differently. There are three
possibilities to remove the small silicate grains selectively: 1) a
composition effect; 2) a size-effect: the silicates are too large to
be seen in that wavelength-region; or 3) aggregates with other
materials. In what follows we will discuss these different effects.

1) A {\bf composition effect}: We expect the strength of the silicate
emission to be related to (among others) the amount of silicates, as
the emitting dust around our sources is optically thin. However, it
is unrealistic to assume that the sources which do not show silicate
emission have no silicates at all in their disc. This would mean that
there were no silicates in the material from which the group Ib stars
are formed. This hypothesis is most unlikely, since the material from
which stars are formed is the ISM, which is relatively uniform in
composition. There is no reason to assume that the silicates were not
there in the beginning around some stars and were there around
others. Besides, the presence or absence of silicate emission does not
depend on the {\em initial} composition: two very young HAEBE stars,
$\mathrm{LkH_{\alpha}}$ 224 and $\mathrm{LkH_{\alpha}}$ 225 are
located close to one another, so must be formed out of the same
material. Surprisingly, the first object does not show any silicate
band, while the second object does show silicate absorption (van den
Ancker 1999). This observation favours a scenario where the presence
or absence of silicate emission is determined by other characteristics
than only the chemical composition of the CS disc.

2) A {\bf size effect}: Since we found evidence for hot grains (from
the 2-10~\mic~excess) capable of producing a strong 10~\mic~feature if
they are (partially) composed of silicates, we must come to the
conclusion that the silicate grains around e.g. HD100453 and HD169142
(both group Ib objects) do not produce a silicate bump because these
grains are on average {\em larger} than the wavelength
(i.e. 10~\mic). Large silicate grains result in an emission resembling
a black body without strong spectral signatures reflecting the
chemical composition. Also Hanner et al. (1994) conclude that the
absence of small silicate grains is the cause for the weak silicate
emission features in comets Austin and Okazaki-Levy-Rudenko.

The absence of the 10~\mic~silicate feature does not necessarily mean
that also the 18~\mic~silicate feature must be absent: e.g. in NGC
6302 (Lim et al. 2000), the 10~\mic~silicate feature is supressed
because of the cold temperature of the silicate dust and the dominant
emission of the C-rich dust, but a 18~\mic~feature is observed.
Therefore, we searched the group Ib sources for silicate features at
18~\mic. But unlike group Ia sources, group Ib sources do not require
solid state bands in addition to the BB+power-law to fit their
continuum. We cannot fully exclude the possibility that small
silicates are present in the discs of group Ib stars, but they must be
so by a much smaller amount and/or colder than in group Ia and IIa
objects.  Sylvester et al. (1996), however, claim to detect silicates
(around 18 \mic) in the spectra of HD169142 and HD135344, indicating
that these stars do have larger silicate particles; from our data, we
cannot confirm this observation, however.  Modelling should determine
how much silicate material can be present in the dust without being
revealed in the spectrum.

%[if T BB $<$175K $->$ dan kunnen we geen silicaten zien!] 
                                                 
3) Another possibility is that, after {\bf coagulation}, small
silicate particles are locked up into larger grains, composed of both
small silicate grains and some other material (as is seen in
interplanetary dust particles (IDPs)). This would make them invisible
if the mantle surrounding the silicate is sufficiently thick. If these
coagulated grains are transported towards the star, the silicate
material will start to evaporate when close enough, while Fe or
carbonaceous material can survive at higher temperatures. This
scenario can account for both the presence of a near-IR excess and for
the absence of hot silicate grains.

\section{Conclusion}

ISO-SWS spectra have shown that there is a large diversity concerning
IR spectral features and shapes in Herbig Ae/Be stars. The results from 
this ISO-sample can be summarized as follows: 

\begin{enumerate}

\item{our sample of 14 HAEBE stars can be classified into two main groups, 
based upon the shape of the continuum (flat or rising); a division which is
further supported by the decomposition of the continuum into a power-law 
and a BB}

\item{the disc geometry is as follows: a geometrically thin disc 
being responsible for the power-law continuum, and a flared region of warm 
dust around the thin disc, causing a rising continuum in the IR. An 
optically thin inner disc causes the solid state bands. Hot dust in the 
inner part of the disc is responsible for the near-IR excess. This disc 
geometry has also been proposed for TTS (Chiang \& Goldreich, 1997)}

\item{the near-IR spectral region is very similar and the sub-mm emission
substantial around all sample sources, indicating homogeneity of the 
hot material and survival of large grains throughout the whole pre-main 
sequence evolution towards the MS}

\item{the mid to far-IR region, on the contrary, is very diverse, and we
attribute this to the amount of flaring in the disc} 

\item{group I sources may evolve into group II sources; the latter
have evidence for larger grains and lack the flared region present in 
group I sources}      

\item{the {\bf presence} of PAH bands cannot be correlated to any of the 
stellar parameters, but they are only present in stars with a large amount 
of warm dust (group I). They are most probably located in an extended region 
around the disc, where they are irradiated by the star. Our PAH bands differ 
from PAH bands in the ISM}

\item{surprising is the {\bf independent behaviour} of the silicate
grains: although other small particles are still present, small
silicate grains seem to be absent around several stars.  This poses an
intriguing problem: what happens to the silicates, what causes them to
behave so differently? It is important to get a definitive answer on
the warm silicates, to know to which extent they can be hidden. Only
then can the observations and proposed model converge to a consistent
picture. Therefore, the next step in our study will be a detailed
modelling of some of the sources}

\end{enumerate}

\acknowledgements{We would like to thank B. Vandenbussche for
assisting with the data reduction and IDL; R. Sylvester, C. Dominik 
and A. de Koter for discussions on CS discs; and S. Hony and 
C. Van Kerckhoven for fruitful discussions about PAHs. GM acknowledges 
financial support from the Flemish Institute for fostering scientific 
and technological research in industry (IWT) under grant IWT/SB/951067. 
LBFMW acknowledges financial support from NWO pionier grant number 
616.078.333.}

\end{document}